\documentclass[12pt]{article}
\usepackage{amssymb}
\usepackage{amsmath}
\usepackage{color}
\usepackage{cancel}
\usepackage[affil-it]{authblk}
 
\usepackage{graphicx}

\newtheorem{theorem}{Theorem}
\newtheorem{prop}[theorem]{Proposition}

\newtheorem{remark}[theorem]{Remark}
\newenvironment{rem}{\begin{remark} \rm}{\end{remark}}

\usepackage{cancel}

\def\parpo#1#2{\{#1,#2\}}

\newcommand{\RR}{{{\mathbb{R}}}}
\newcommand{\brke}[2]{{{\langle #1,#2\rangle}}}

\def\M{{\mathcal M}}
\def\Tr{\operatorname{Tr}}

\begin{document}
\title{Poisson quasi-Nijenhuis manifolds \\ and the Toda system}

\author{G. Falqui${}^1$, I. Mencattini${}^2$, G. Ortenzi${}^1$, M. Pedroni${}^3$}

\affil{
{\small  $^1$Dipartimento di Matematica e Applicazioni, 
Universit\`a di Milano-Bicocca, 
Italy
}\\
{\small  gregorio.falqui@unimib.it (ORCID 0000-0002-4893-9186)}\\
{\small giovanni.ortenzi@unimib.it (ORCID 0000-0003-2192-6737)}\\
\medskip
{\small $^2$Instituto de Ci\^encias Matem\'aticas e de Computa\c c\~ao,  Universidade de S\~ao Paulo, Brazil}\\
{\small igorre@icmc.usp.br (ORCID 0000-0002-1295-9396)}\\
\medskip
{\small $^3$Dipartimento di Ingegneria Gestionale, dell'Informazione e della Produzione, 
Universit\`a di Bergamo, 
Italy}\\
{\small marco.pedroni@unibg.it (corresponding author, ORCID 0000-0002-7358-0945)}
}

\maketitle
\abstract{\noindent 
The notion of Poisson quasi-Nijenhuis manifold generalizes that of Poisson-Nijenhuis manifold. The relevance of the latter in the theory of completely integrable systems is well established since the birth of the bi-Hamiltonian approach to integrability. In this note, we discuss the relevance of the notion of Poisson quasi-Nijenhuis manifold in the context of finite-dimensional integrable systems. Generically (as we show by an example with $3$ degrees of freedom) the Poisson quasi-Nijenhuis structure is largely too general to ensure Liouville integrability of a system. However, we prove that the closed (or periodic) $n$-particle Toda lattice can be framed in such a geometrical structure, and its well-known integrals of the motion can be obtained as spectral invariants of a ``quasi-Nijenhuis recursion operator'', that is, a tensor field $N$ of type $(1,1)$ defined on the phase space of the lattice. This example and some of its generalizations are used to understand whether one can define in a reasonable sense a notion of {\em involutive\/} Poisson quasi-Nijenhuis manifold. A geometrical link between the open (or non periodic) and the closed Toda systems is also framed in the context of a general scheme connecting Poisson quasi-Nijenhuis and Poisson-Nijenhuis manifolds.
}  

\baselineskip=0,6cm

\section{Introduction}
It is well known that Poisson-Nijenhuis (PN) manifolds \cite{MagriMorosiRagnisco85,KM} are an important notion in the theory of integrable systems. Roughly speaking, they are Poisson manifold $(\M,\pi)$ endowed with a tensor field of type $(1,1)$, say $N:T\M\to T\M$, which is torsionless and compatible (see Section \ref{subsec:traces}) with the Poisson tensor $\pi$. They turn out to be bi-Hamiltonian manifolds, with the traces of the powers of $N$ satisfying the Lenard-Magri relations and thus being in involution with respect to the Poisson brackets induced by the Poisson tensors. An example of integrable system that can be studied in the context of PN manifolds is the open (or non periodic) $n$-particle Toda lattice. (For both the periodic and the non periodic Toda system, see \cite{Perelomov-book} and references therein; see also \cite{Damianou04,MorosiPizzocchero96,Okubo90}.) 
The PN structure of the open Toda lattice was presented in \cite{DO}. Its Poisson tensor is non degenerate, so that the PN manifold is a symplectic manifold (sometimes it is called an $\omega$N-manifold). This kind of geometrical structure was shown to play an important role in the bi-Hamiltonian interpretation of the separation of variable method (see, e.g., \cite{FMP01,FP03}).

Poisson quasi-Nijenhuis (PqN) manifolds are an interesting generalization of PN manifolds. They were introduced in \cite{SX}, where the requirement about the vanishing of the (Nijenhuis) torsion of $N$ is weakened in a suitable sense, and the relations with quasi-Lie bialgebroid and symplectic Nijenhuis groupoids are investigated. In their Remark 3.13, the authors write: ``Poisson Nijenhuis structures arise naturally in the study of integrable systems. It would be interesting to find applications of Poisson quasi-Nijenhuis structures in integrable systems as well."
As far as we know, no progress in this direction was made until now. 

The aim of this paper is to interpret the well known integrability of the closed Toda lattice in the framework of PqN manifolds. More precisely, we introduce a tensor field $N$ of type $(1,1)$ which is compatible with the canonical Poisson tensor and endows $\RR^{2n}$ with the structure of a PqN manifold, and we show that the traces $I_k$ of the powers of $N$ are integrals of motion in involution. However, we discuss a class 
of PqN manifolds clarifying that the involutivity of the $I_k$ does not hold in every PqN manifold. 

The organization of this paper is the following. In Section \ref{subsec:traces} we recall the definitions of PN and PqN manifold, and we show how the classical Lenard-Magri recursion relations among the $I_k$ are modified in the PqN case.  Section \ref{laclassediesempidiGiovanni} is devoted to a class of PqN structures on $\RR^6$ depending on a potential $V$ and showing that the $I_k$ are in involution only for special choices of $V$. In Section \ref{toda-4} we consider the 4-particle closed Toda system with its PqN structure, performing some computations on the $I_k$ to prove that they are in involution. These results are generalized in Section \ref{FS}, while in 
Section \ref{igor-connection} we present general results clarifying the relation between the PN structure of the open Toda lattice and the PqN structure of the closed one.


\par\smallskip\noindent
{\bf Acknowledgments.}
We wish to thank Yvette Kosmann-Schwarzbach, Franco Magri and, especially, Orlando Ragnisco for useful discussions. 
MP thanks  the {Dipartimento di Matematica e Applicazioni} of 
Universit\`a Milano-Bicocca  for its hospitality. 
This project has received funding from the European Union's Horizon 2020 research and innovation programme under the Marie Sk{\l}odowska-Curie grant no 778010 {\em IPaDEGAN}. All authors gratefully acknowledge the auspices of the GNFM Section of INdAM under which part of this work was carried out.


\section{Nijenhuis torsion and Poisson quasi-Nijenhuis manifolds}
\label{subsec:traces}
It is well known that the {\it Nijenhuis torsion\/} of a $(1,1)$ tensor field $N:T\M\to T\M$ on a manifold $\M$ is defined as 
\begin{equation}
\label{tndef1}
T_N(X,Y)=[NX,NY]-N\left([NX,Y]+[X,NY]-N[X,Y]\right)\, .
\end{equation}
It can be written as 
\begin{equation}
\label{tndef2}
T_N(X,Y)=(L_{NX}N-NL_XN)
 Y\,, 
\end{equation}
where, hereafter, $L_X$ denotes the Lie derivative with respect to the vector field $X$. Hence one arrives at the formula
\begin{equation}
\label{ff1}
N\, L_X N=L_{NX}N-i_XT_N\, ,
\end{equation}
where $i_XT_N$ is the $(1,1)$ tensor field obviously defined as $(i_XT_N)(Y)=T_N(X,Y)$. We recall that, given a $p$-form $\alpha$, with $p\ge 1$, one can construct another $p$-form $i_N\alpha$ as 
\begin{equation}
\label{iNalpha}
i_N\alpha(X_1,\dots,X_p)=\sum_{i=1}^p \alpha(X_1,\dots,NX_i,\dots,X_p)\,,
\end{equation}
and that $i_N$ is a derivation 
of degree zero (if $i_N f=0$ for all function $f$).
We also remind \cite{MagriMorosiRagnisco85} that $N:T\M\to T\M$ and a Poisson bivector $\pi:T^*\M\to T\M$ are said to be {\it compatible\/} if 
\begin{equation}
\label{N-P-compatible}
\begin{split}
&N\pi=\pi N^*\,,\qquad
\mbox{where $N^*:T^*\M\to T^*\M$ is the transpose of $N$;}\\
&L_{\pi\alpha}(N) X-\pi L_{X}(N^*\alpha)+\pi L_{NX}\alpha=0\,,\qquad\mbox{for all 1-forms $\alpha$ and vector fields $X$.}
\end{split}
\end{equation}
Some nice interpretations of these compatibility conditions 
were given in \cite{YKS96}. We will use one of them in Section \ref{igor-connection}.

In \cite{SX} a {\it Poisson quasi-Nijenhuis (PqN) manifold\/} was defined as a quadruple $(\M,\pi,N, \phi)$ such that 
\begin{itemize}
\item the Poisson bivector $\pi$ and the $(1,1)$ tensor field $N$  
are compatible;
\item the 3-forms $\phi$ and $i_N\phi$ are closed;
\item $T_N(X,Y)=\pi\left(i_{X\wedge Y}\phi\right)$ for all vector fields $X$ and $Y$, where $i_{X\wedge Y}\phi$ is the 1-form defined as 
$\langle i_{X\wedge Y}\phi,Z\rangle=\phi(X,Y,Z)$.
\end{itemize}
The bivector field $\pi'=N\pi$ turns out to satisfy the conditions 
\begin{equation}
\label{schouten-conditions}
[\pi,\pi']=0\,,\qquad [\pi',\pi']=2\pi(\phi)\, ,
\end{equation}
where $[\cdot,\cdot]$ is the Schouten bracket (see, e.g., \cite{Izu}) between bivectors and 
$\pi(\phi)(\alpha,\beta,\gamma)=\phi(\pi\alpha,\pi\beta,\pi\gamma)$ 
for any triple of 1-forms $(\alpha,\beta,\gamma)$.
The following result, also proved in \cite{SX}, will be used in this paper.
\begin{prop}
\label{prop:piprime}
Let $\M$ be a manifold endowed with a non degenerate Poisson tensor $\pi$, a tensor field $N$ of type $(1,1)$, and a closed 3-form $\phi$. 
If $N\pi=\pi N^*$ and conditions (\ref{schouten-conditions}) are satisfied (with $\pi'=N\pi$), then $(\M,\pi,N, \phi)$ is a PqN manifold. 
\end{prop}

If $\phi=0$, then the torsion of $N$ vanishes and $\M$ becomes a {\it Poisson-Nijenhuis manifold} (see \cite{KM} and references therein). The bivector field $\pi'=N\pi$ is in this case a Poisson tensor compatible with $\pi$. 
Moreover, the functions
\begin{equation}
\label{tracce}
I_k=\frac1{k}\Tr(N^k)\,,\qquad k=1,2,\dots\, ,
\end{equation}
satisfy $dI_{k+1}=N^* dI_{k}$, entailing the so-called {\it Lenard-Magri relations\/} 
\begin{equation}
\label{LM-rel}
\pi dI_{k+1}=\pi' dI_{k}
\end{equation}
and therefore the involutivity of the $I_k$ (with respect to both Poisson brackets induced by $\pi$ and $\pi'$). 

For a general PqN manifold $\M$, we will see in the next section  
that such involutivity (with respect to the unique Poisson bracket defined on $\M$, i.e., the one associated with $\pi$) does not hold. Anyway, we have that, for $k\ge 2$ and for a generic vector field $X$ on $\M$, 
\begin{equation}
\label{req1}
\begin{split}
\brke{dI_{k+1}}{X}&=L_X\left(\frac{1}{k+1} \Tr(N^{k+1})\right)=\Tr\left((NL_X N)N^{k-1}\right)\\
&\buildrel(\ref{ff1})\over=\Tr\left(L_{NX}(N)N^{k-1}\right)-\Tr\left((i_XT_N)\, N^{k-1}\right)\\ 
&=L_{NX}\left(\frac1{k} \Tr(N^{k})\right)-\Tr\left((i_XT_N)\, N^{k-1}\right)\\ &=\brke{dI_{k}}{NX}-\Tr\left((i_XT_N)\, N^{k-1}\right)\\ 
&=\brke{N^*dI_{k}}{X}-\Tr\left((i_XT_N)\, N^{k-1}\right)\, .
\end{split}
\end{equation}
So we arrive at the generalized Lenard-Magri relations 
\begin{equation}
\label{ff2}
dI_{k+1}=N^* dI_{k}-{\phi_{k-1}}
\, ,
\end{equation}
where we used the definition 
\begin{equation}
\label{varphi}
\brke{\phi_l}{ X}= \Tr\left((i_XT_N)\, N^{l}\right)= \Tr\left(N^{l}\, (i_XT_N)\right)\, ,\qquad l\ge 0\, .
\end{equation}
Notice that this definition, along with (\ref{ff2}), was used in \cite{Bogo96-180,Bogo96-182} for different purposes. 
Let us compute now the Poisson bracket $\parpo{I_k}{I_j}$ for $k>j\ge 1$:
\begin{equation}
\label{rectr}
\begin{split}
\parpo{I_k}{I_j}&=\brke{dI_k}{\pi dI_j}\buildrel{(\ref{ff2})}\over=\brke{N^*d I_{k-1}}{ \pi dI_j}-\brke{\phi_{k-2}}{ \pi dI_j}=\brke{dI_{k-1}}{N\pi dI_j}-\brke{\phi_{k-2}}{ \pi dI_j}\\ &=\brke{dI_{k-1}}{\pi\, N^* dI_{j}}-\brke{\phi_{k-2}}{ \pi dI_j}\buildrel{(\ref{ff2})}\over=
\brke{dI_{k-1}}{\pi dI_{j+1}}+\brke{dI_{k-1}}{\pi \phi_{j-1}}-\brke{\phi_{k-2}}{ \pi dI_j}
\\ \noalign{\medskip}
&=\parpo{I_{k-1}}{I_{j+1}}-\left(\brke{ \phi_{j-1}}{\pi dI_{k-1}}+\brke{\phi_{k-2}}{ \pi dI_j}\right)\, .
\end{split}
\end{equation}
Thus, the usual formula  
\begin{equation}
\label{LMrec}
\parpo{I_k}{I_j}=\parpo{I_{k-1}}{I_{j+1}}\,,
\end{equation}
entailed by the Lenard-Magri relations (\ref{LM-rel}), in the non vanishing torsion case is modified as follows:
\begin{equation}
\label{recadd}
\parpo{I_k}{I_j}-\parpo{I_{k-1}}{I_{j+1}}=-\left(\brke{ \phi_{j-1}}{\pi dI_{k-1}}+\brke{\phi_{k-2}}{ \pi dI_j}\right)\, .
\end{equation}
Actually, one can see that the 1-forms $\phi_l$ compute the Poisson brackets between the $I_j$. 
Indeed, if we consider $k=j+1$, we obtain from
(\ref{recadd})
\begin{equation}
\label{exk=j+1}
\parpo{I_{j+1}}{I_j}=-\brke{ \phi_{j-1}}{\pi dI_{j}}\, .
\end{equation}
A necessary condition for the traces of the powers of $N$ to be in involution is thus $\brke{ \phi_{j-1}}{\pi dI_{j}}=0$ for all $j\ge 1$, 
which explicitly reads
\begin{equation}\label{jj+1}
\Tr\left((i_{\pi dI_j} T_N)\, N^{j-1}\right)=0\, .
\end{equation}
However, imposing the condition that 
\begin{equation}
\label{nocond}
\brke{ \phi_{k}}{\pi dI_{j}}=
\Tr\left((i_{\pi dI_j} T_N)\, N^{k}\right)=0
\end{equation}
for all $k,j$  (although being clearly 
sufficient), is too restrictive: indeed, it fails in the simplest non trivial 
case, namely, the closed Toda system with $4$ particles (see Section \ref{Toda4}).

Some further conditions can be written, which explain the above sentence in general. For example, if we take $k=j+2$ we obtain, still from 
(\ref{recadd}),
\begin{equation}
\label{exk=j+2}
\parpo{I_{j+2}}{I_j}={\parpo{I_{j+1}}{I_{j+1}}}-\left(\brke{ \phi_{j-1}}{\pi dI_{j+1}}+\brke{\phi_{j}}{ \pi dI_j}\right)\, .
\end{equation}
To ensure that $\parpo{I_{j+2}}{I_j}$ be zero, no need that the  last two   terms in the right-hand side of the above equation be simultaneously vanishing. Indeed,  the Toda closed chain with $4$ particles is already an example in which these two terms {\em cancel\/} each other without  vanishing on their own.


\section{A class of non involutive PqN manifolds}
\label{laclassediesempidiGiovanni} 

In this section we present a wide class of examples of PqN manifolds such that the traces (\ref{tracce}) are not in involution. Let us consider, on $\M=\RR^6$ with (canonical) variables $(q_{{1}},q_{{2}},q_{{3}},p_{{1}},p_{{2}},p_{{3}})$, the canonical Poisson tensor $\pi$ and the $(1,1)$ tensor field given by
\begin{equation}
\label{qN3-V}
N=\left[ \begin {array}{cccccc} 
p_{{1}}&0&0&0&1&1\\ \noalign{\medskip}
0&p_{{2}}&0&-1&0&1\\ \noalign{\medskip}
0&0&p_{{3}}&-1&-1&0\\ \noalign{\medskip}
0&-V(q_1-q_2)&-V(q_3-q_1)&p_{{1}}&0&0\\ \noalign{\medskip}
V(q_1-q_2)&0&-V(q_2-q_3)&0&p_{{2}}&0\\ \noalign{\medskip}
V(q_3-q_1)&V(q_2-q_3)&0&0&0&p_{{3}}
\end {array} \right]\,,
\end{equation}
where $V$ is an arbitrary (differentiable) function of one variable.
First of all, we use Proposition \ref{prop:piprime} to show that $\pi$ and $N$ define, together with a suitable 3-form $\phi$, a PqN structure on 
$\RR^6$. Indeed, if
\begin{equation}
\label{P1-V}
\pi'=N\pi=\left[ \begin {array}{cccccc} 
0&-1&-1&p_{{1}}&0&0\\ \noalign{\medskip}
1&0&-1&0&p_{{2}}&0\\ \noalign{\medskip}
1&1&0&0&0&p_{{3}}\\ \noalign{\medskip}
-p_{{1}}&0&0&0&-V(q_1-q_2)&-V(q_3-q_1)\\ \noalign{\medskip}
0&-p_{{2}}&0&V(q_1-q_2)&0&-V(q_2-q_3)\\ \noalign{\medskip}
0&0&-p_{{3}}&V(q_3-q_1)&V(q_2-q_3)&0
\end {array}\right]\,,
\end{equation}
then one can easily show that $[\pi,\pi']=0$, so that the first of (\ref{schouten-conditions}) holds. Moreover, we have that 
\begin{equation}
\label{Scbr-V}
\begin{split}
[\pi',\pi']&=2\left(V(q_1-q_2)-V'(q_1-q_2)\right)\partial_{p_2}\wedge \partial_{p_1}\wedge \left(\partial_{q_1}+\partial_{q_2}\right)\\
&+2\left(V(q_2-q_3)-V'(q_2-q_3)\right)\partial_{p_3}\wedge \partial_{p_2}\wedge \left(\partial_{q_2}+\partial_{q_3}\right)\\
&+2\left(V(q_3-q_1)+V'(q_3-q_1)\right)\partial_{p_3}\wedge \partial_{p_1}\wedge \left(\partial_{q_1}+\partial_{q_3}\right)\\
&+4 V'(q_3-q_1)\partial_{p_3}\wedge \partial_{p_1}\wedge \partial_{q_2}\,.
\end{split}
\end{equation}
Therefore, the 3-form $\phi$ such that $[\pi',\pi']=2\pi(\phi)$ turns out to be
\begin{equation}
\label{Phi-V}
\begin{split}
\phi &=\left(V'(q_1-q_2)-V(q_1-q_2)\right)d(p_1+p_2)\wedge d{q_2}\wedge d{q_1}\\
&+\left(V'(q_2-q_3)-V(q_2-q_3)\right)d(p_2+p_3)\wedge d{q_3}\wedge d{q_2}\\
&-\left(V'(q_3-q_1)+V(q_3-q_1)\right)d(p_1+p_3)\wedge d{q_3}\wedge d{q_1}\\
&-2 V'(q_3-q_1)dp_2\wedge d{q_3}\wedge d{q_1}\,,
\end{split}
\end{equation}
which is clearly closed. Hence we can conclude by Proposition \ref{prop:piprime} that $(\RR^6, \pi, N, \phi)$ is a PqN manifold for every choice of the function $V$.
One can check that $T_N(X,Y)=\pi\left(i_{X\wedge Y}\phi\right)$ for all vector fields $X,Y$, as stated in \cite{SX}.

Consider now the functions $H_k=\frac12 I_k=\frac1{2k}\Tr(N^k)$. We have that $H_1=p_1+p_2+p_3$, 
\begin{equation}
\label{ham-V}
H_2=\frac12({p_1}^2+{p_2}^2+{p_3}^2)+V(q_1-q_2)+V(q_2-q_3)+V(q_3-q_1)\,,
\end{equation}
which can be obviously thought of as the Hamiltonian of three interacting particles of equal mass, and 
\begin{equation}
\label{ham3-V}
\begin{split}
H_3&=\frac13({p_1}^3+{p_2}^3+{p_3}^3)+p_1\left(V(q_1-q_2)+V(q_3-q_1)\right)
+p_2\left(V(q_2-q_3)+V(q_1-q_2)\right)\\
&+p_3\left(V(q_3-q_1)+V(q_2-q_3)\right)\,.
\end{split}
\end{equation}
It is clear that $\parpo{H_1}{H_2}=\parpo{H_1}{H_3}=0$, while the Poisson bracket
\begin{equation}
\label{invH2H3-V}
\begin{split}
\parpo{H_2}{H_3}&=V(q_1-q_2) \left(V'(q_2-q_3)-V'(q_3-q_1)\right)
+V(q_2-q_3)  \left(V'(q_3-q_1)-V'(q_1-q_2)\right)\\
&+V(q_3-q_1)  \left(V'(q_1-q_2)-V'(q_2-q_3)\right)\,
\end{split}
\end{equation}
does not vanish for any function $V$ (for example, one can easily check that it is different from zero if $V(x)=1/x$). However, involutivity holds in the cases
$V(x)=\mathrm{e}^x$ (to be discussed in the next sections) and $V(x)=1/x^2$ (corresponding to the Calogero model). 

In conclusion, given a PqN manifold, further conditions on $(\pi,N,\phi)$ are needed to guarantee that the functions $I_k$ are in involution.


\section{The 4-particle closed Toda case}\label{toda-4}
In this section we consider the closed (or periodic) Toda system with $n=4$ particles. In the canonical variables 
$(q_{{1}},q_{{2}},q_{{3}},q_{{4}},p_{{1}},p_{{2}},p_{{3}},p_{{4}})$,
the Hamiltonian is given by 
\begin{equation}
\label{Htoda}
H_{\mbox{\scriptsize Toda}}=\sum_{i=1}^4 \left(\frac12 p_i^2+\mathrm{e}^{q_i-q_{i+1}}\right),\qquad\mbox{where $q_{5}
=q_1$.}
\end{equation}
Let us introduce the $(1,1)$
tensor field on $\M=\RR^8$ given by
\begin{equation}
\label{qN4}
N=\left[ \begin {array}{cccccccc} p_{{1}}&0&0&0&0&1&1&1
\\ \noalign{\medskip}0&p_{{2}}&0&0&-1&0&1&1\\ \noalign{\medskip}0&0&p_
{{3}}&0&-1&-1&0&1\\ \noalign{\medskip}0&0&0&p_{{4}}&-1&-1&-1&0
\\ \noalign{\medskip}0&-{{\rm e}^{q_{{1}}-q_{{2}}}}&0&-{{\rm e}^{q_{{4}
}-q_{{1}}}}&p_{{1}}&0&0&0\\ \noalign{\medskip}{{\rm e}^{q_{{1}}-q_{{2}
}}}&0&-{{\rm e}^{q_{{2}}-q_{{3}}}}&0&0&p_{{2}}&0&0
\\ \noalign{\medskip}0&{{\rm e}^{q_{{2}}-q_{{3}}}}&0&-{{\rm e}^{q_{{3}
}-q_{{4}}}}&0&0&p_{{3}}&0\\ \noalign{\medskip}{{\rm e}^{q_{{4}}-q_{{1
}}}}&0&{{\rm e}^{q_{{3}}-q_{{4}}}}&0&0&0&0&p_{{4}}\end {array}
 \right]\,
\end{equation}
and the traces of its powers, $I_k=\frac1{k}\Tr(N^k)$.
As we will see, these functions are in involution with respect to the (canonical) Poisson bracket $\parpo{\cdot}{\cdot}$ induced by the 
canonical Poisson tensor $\pi$. If we put $H_k=\frac12 I_k$, then it is easy to check that $H_1=\sum_{i=1}^4 p_i$ and $H_2=H_{\mbox{\scriptsize Toda}}$,  while $H_3$ is the third constant of the motion of the $4$-particle Toda chain, and $H_4$ coincides with the fourth one up to a constant. Here, by ``constants of the motion of the $4$-particle Toda chain'' we mean those obtained by taking traces of the powers of the well known Lax matrix (see, e.g.,  \cite{Perelomov-book})
\begin{equation}
L=\left[
\begin{array}{cccc}\medskip
p_1&{{\rm e}^{\frac12(q_{{1}}-q_{{2}) }}}&0&{{\rm e}^{\frac12(q_{{4}}-q_{{1}) }}}\\ \medskip
{{\rm e}^{\frac12(q_{{1}}-q_{{2}) }}}&p_2&{{\rm e}^{\frac12(q_{{2}}-q_{{3}) }}}&0\\ \medskip
0&{{\rm e}^{\frac12(q_{{2}}-q_{{3}) }}}&p_3&{{\rm e}^{\frac12(q_{{3}}-q_{{4}) }}}\\ \medskip
{{\rm e}^{\frac12(q_{{4}}-q_{{1}) }}}&0&{{\rm e}^{\frac12(q_{{3}}-q_{{4}) }}}&p_4
\end{array}\right]\>.
\end{equation}

%
We can use Proposition \ref{prop:piprime} to show that $\pi$ and $N$ define a PqN structure on $\RR^8$. 
Indeed, $N$ differs from the {\em torsion free\/} $(1,1)$ tensor field of the {\em open\/} Toda chain (see \cite{DO}), 
\begin{equation}
\label{Nopen}
N_{(O)}=\left[ \begin {array}{cccccccc} p_{{1}}&0&0&0&0&1&1&1
\\ \noalign{\medskip}0&p_{{2}}&0&0&-1&0&1&1\\ \noalign{\medskip}0&0&p_
{{3}}&0&-1&-1&0&1\\ \noalign{\medskip}0&0&0&p_{{4}}&-1&-1&-1&0
\\ \noalign{\medskip}0&-{{\rm e}^{q_{{1}}-q_{{2}}}}&0&0&p_{{1}}&0&0&0
\\ \noalign{\medskip}{{\rm e}^{q_{{1}}-q_{{2}}}}&0&-{{\rm e}^{q_{{2}}-
q_{{3}}}}&0&0&p_{{2}}&0&0\\ \noalign{\medskip}0&{{\rm e}^{q_{{2}}-q_{{
3}}}}&0&-{{\rm e}^{q_{{3}}-q_{{4}}}}&0&0&p_{{3}}&0
\\ \noalign{\medskip}0&0&{{\rm e}^{q_{{3}}-q_{{4}}}}&0&0&0&0&p_{{4}}
\end {array} \right]\,,
\end{equation}
by the rank $2$ tensor
\begin{equation}
\label{delta}
\Delta N=\mathrm{e}^{q_4-q_1}\left(\partial_{p_4}\otimes dq_1-\partial_{p_1}\otimes dq_4\right)\, .
\end{equation}
It can be checked that the torsion of $\Delta N$ vanishes, while that 
of $N$ turns out to be
\begin{equation}
\label{TN4}
T_N=\mathrm{e}^{q_4-q_1}\left(\partial_{p_1}\otimes dq_4\wedge 
dI_1
-\partial_{p_4}\otimes dq_1\wedge 
dI_1
-X_1\otimes dq_1\wedge dq_4\right)\, ,
\end{equation}
where $X_1=\pi dI_1=2\sum_{i=1}^4\partial_{q_i}$ is (twice) the translation vector field. In other words, 
\begin{equation}
\label{Igor-TN}
T_N(X,Y)=\langle dI_1,Y\rangle\Delta N(X)-\langle dI_1,X\rangle\Delta N(Y)+\Omega(X,Y)X_1,
\end{equation}
where $\Omega=\mathrm{e}^{q_4-q_1}dq_4\wedge dq_1$. It is easily seen that (\ref{Igor-TN}) holds for the general $n$-particle case, with 
$\Omega=\mathrm{e}^{q_n-q_1}dq_n\wedge dq_1$.

We also have that 
\begin{equation}
\label{P1}
\pi'=N\pi=\left[ \begin {array}{cccccccc} 0&-1&-1&-1&p_{{1}}&0&0&0
\\ \noalign{\medskip}1&0&-1&-1&0&p_{{2}}&0&0\\ \noalign{\medskip}1&1&0
&-1&0&0&p_{{3}}&0\\ \noalign{\medskip}1&1&1&0&0&0&0&p_{{4}}
\\ \noalign{\medskip}-p_{{1}}&0&0&0&0&-{{\rm e}^{q_{{1}}-q_{{2}}}}&0&
-{{\rm e}^{q_{{4}}-q_{{1}}}}\\ \noalign{\medskip}0&-p_{{2}}&0&0&{{\rm e}
^{q_{{1}}-q_{{2}}}}&0&-{{\rm e}^{q_{{2}}-q_{{3}}}}&0
\\ \noalign{\medskip}0&0&-p_{{3}}&0&0&{{\rm e}^{q_{{2}}-q_{{3}}}}&0&-{
{\rm e}^{q_{{3}}-q_{{4}}}}\\ \noalign{\medskip}0&0&0&-p_{{4}}&{
{\rm e}^{q_{{4}}-q_{{1}}}}&0&{{\rm e}^{q_{{3}}-q_{{4}}}}&0\end {array}
 \right]\,,
\end{equation}
while the corresponding {\em Poisson\/} tensor for the open Toda lattice is
\begin{equation}
\label{P1O}
\pi'_{(O)}=N_{(O)}\pi= \left[ \begin {array}{cccccccc} 0&-1&-1&-1&p_{{1}}&0&0&0
\\ \noalign{\medskip}1&0&-1&-1&0&p_{{2}}&0&0\\ \noalign{\medskip}1&1&0
&-1&0&0&p_{{3}}&0\\ \noalign{\medskip}1&1&1&0&0&0&0&p_{{4}}
\\ \noalign{\medskip}-p_{{1}}&0&0&0&0&-{{\rm e}^{q_{{1}}-q_{{2}}}}&0&0
\\ \noalign{\medskip}0&-p_{{2}}&0&0&{{\rm e}^{q_{{1}}-q_{{2}}}}&0&-{
{\rm e}^{q_{{2}}-q_{{3}}}}&0\\ \noalign{\medskip}0&0&-p_{{3}}&0&0&{
{\rm e}^{q_{{2}}-q_{{3}}}}&0&-{{\rm e}^{q_{{3}}-q_{{4}}}}
\\ \noalign{\medskip}0&0&0&-p_{{4}}&0&0&{{\rm e}^{q_{{3}}-q_{{4}}}}&0
\end {array} \right]\,.
\end{equation}
It holds
\begin{equation}
\label{delta P1}
\pi'=\pi'_{(O)}+\mathrm{e}^{q_4-q_1}
\partial_{p_4}\wedge\partial_{p_1}\,,
\end{equation}
and the Schouten bracket of $\pi'$ with itself is
\begin{equation}
\label{Scbr}
[\pi',\pi']
=2\mathrm{e}^{q_4-q_1}\left(X_1\wedge \partial_{p_4}\wedge \partial_{p_1}\right)\,.
\end{equation}
Then we find that the second of (\ref{schouten-conditions}) is satisfied with
\begin{equation}
\label{Phi}
\phi=\mathrm{e}^{q_4-q_1}\left(dI_1\wedge dq_1\wedge dq_4\right)
= dI_1\wedge d \mathrm{e}^{q_4}\wedge d\mathrm{e}^{-q_1}=d \left( I_1\, d \mathrm{e}^{q_4}\wedge d\mathrm{e}^{-q_1}\right)\, ,
\end{equation}
which is obviously closed. Hence one is left with showing that the first of (\ref{schouten-conditions}) holds, which is a quite easy task. Using Proposition \ref{prop:piprime}, we can then conclude that $(\RR^8, \pi, N, \phi)$ is a PqN manifold.

\begin{rem}
As we have already seen in Section \ref{subsec:traces}, many features of the usual picture of Poisson-Nijenhuis manifolds are lost in the case. Not only $\pi'$ is not Poisson, but the Hamiltonians $I_k$ do not fulfill the Lenard-Magri relations. For example, $N^*dI_1\not=dI_2$, so that $N^*dH_1\not=dH_2$ and 
$N X_1\not=X_2$, where $X_2=\pi dI_2$ is twice the ``physical" Toda vector field $X_{\mbox{\scriptsize Toda}}=\pi dH_2$. However, we will show  
that the $I_k$ are in involution. This is not true for an arbitrary PqN manifold, as we have seen in Section \ref{laclassediesempidiGiovanni}.
\end{rem}

The generalization of these patterns to the $n$-particle case is clear. In particular, to obtain the corresponding formulas of 
(\ref{delta}, \ref{TN4}, \ref{delta P1}, \ref{Scbr}, 
\ref{Phi}), one simply has to make the replacement $4 \mapsto n$.


\subsection{Some computations on the traces}\label{Toda4}
This subsection is devoted to some computations, still for the case $n=4$. First of all, we rewrite formula (\ref{Igor-TN}) as 
\begin{equation}
i_XT_N=\Delta N (X)\otimes dI_1-\langle dI_1,X\rangle\Delta N+X_1\otimes i_X\Omega\,,\label{eq:5}
\end{equation}
where $\Delta N$ is given by (\ref{delta}) and $\Omega=\mathrm{e}^{q_4-q_1}dq_4\wedge dq_1$. We notice that 
$\Delta N=-\pi\,\Omega^\flat$, where $ \Omega^\flat:T\M\to T^*\M$ is defined as usual by $\Omega^\flat(X)=i_X\Omega$.
If we call
$X_j=\pi\, dI_j$ the vector fields of the hierarchy, then we have that $\langle dI_1,X_j\rangle=-\brke{dI_j}{X_1}=0$, since $N$ and hence its traces depend only on the differences $q_{i}-q_{i+1}$ of the coordinates. Therefore
\begin{equation}
\label{eq:5b}
i_{X_j}T_N=\Delta N (X_j)\otimes dI_1+X_1\otimes i_{X_j}\Omega\,,
\end{equation}
so that 
\begin{equation}
\label{eq-pre-tr}
\begin{split}
\brke{\phi_k}{X_j} &=\Tr\left(N^k(i_{X_j}T_N)\right)=\Tr\left(N^k(\Delta N (X_j)\otimes dI_1+X_1\otimes i_{X_j}\Omega)\right)\\
&=\Tr\left((N^k\Delta N )(X_j)\otimes dI_1\right)+\Tr\left((N^k X_1)\otimes i_{X_j}\Omega)\right)
 \,. 
\end{split}
\end{equation}
Both summands coincide with $\Omega(X_j,N^k X_1)$. This is easily seen for the second summand, 
since $\Tr(X\otimes\alpha)=\brke{\alpha}{X}$ for all vector fields $X$ 
and 1-forms $\alpha$. As far as the first one is concerned, 
\begin{equation}
\label{eqtr-b}
\begin{split}
\Tr\left((N^k\Delta N )(X_j)\otimes dI_1\right) &=\brke{dI_1}{(N^k\Delta N )(X_j)}
=-\brke{dI_1}{(N^k\pi\, \Omega^\flat)(X_j)}
=-\brke{dI_1}{(\pi {N^*}^k \Omega^\flat)(X_j)}\\
&=\brke{({N^*}^k \Omega^\flat)(X_j)}{X_1}
=\brke{\Omega^\flat(X_j)}{N^k X_1}
=\Omega(X_j,N^k X_1)\,. 
\end{split}
\end{equation}
Therefore we have obtained the final formula
\begin{equation}
\label{eqtr}
\brke{\phi_k}{X_j}=2\Omega(X_j,N^k X_1)
=2\mathrm{e}^{q_4-q_1} \left(\brke{dq_4}{X_j} \brke{dq_1}{N^kX_1}-\brke{dq_1}{X_j} \brke{dq_4}{N^kX_1}\right)
\,. 
\end{equation}
We are now ready for the explicit computations of the Poisson brackets between the $I_j$. We have just seen that $\parpo{I_1}{I_j}=\brke{dI_1}{X_j}=0$ for all $j$, 
therefore we have to check that 
\begin{equation}\label{3eq}
\parpo{I_3}{I_2}=0\,,\qquad \parpo{I_4}{I_3}=0\,,\qquad \parpo{I_4}{I_2}=0\, .
\end{equation}
Taking 
(\ref{exk=j+1}), (\ref{exk=j+2}) and (\ref{eqtr}) into account, these three relations translate respectively into:
\begin{eqnarray}
\label{neq3-1} &&\brke{\phi_1}{X_2}=2\Omega(X_2,N X_1)=
0\\
\label{neq3-2} &&\brke{\phi_2}{X_3}=2\Omega(X_3,N^2 X_1)=
0\\
\label{neq3-3} &&\brke{\phi_1}{X_3}+\brke{\phi_2}{X_2}=2\Omega(X_3,N X_1)+2\Omega(X_2,N^2 X_1)=
0\,.
\end{eqnarray}
If we call $Y_k=N^{k-1}X_1-X_k$, it turns out (see next section) that 
\begin{equation}\label{Y14eq}
i_{Y_k}\Omega=0\,,\qquad\mbox{that is,}\qquad \brke{dq_1}{Y_k}=\brke{dq_4}{Y_k}=0\, .
\end{equation}
Hence we can show that (\ref{neq3-1}) holds by replacing $NX_1$ with $X_2$. Similarly, (\ref{neq3-2}) reduces to $\Omega(X_3,X_3)=0$, 
so we have shown that $\parpo{I_3}{I_2}=\parpo{I_4}{I_3}=0$.
We are left with $\parpo{I_4}{I_2}$, that is, with (\ref{neq3-3}).
In the light of (\ref{Y14eq}), this can be written as
\begin{equation}
\label{I2I4n-bis}
\Omega(X_3,X_2)+\Omega(X_2,X_3)=0\,,
\end{equation}
which clearly holds. Notice however that, e.g., 
$$
\Tr\left(N\, (i_{X_3}T_N)\right)=\brke{\phi_1}{X_3}=2\Omega(X_3,NX_1)=2\Omega(X_3,X_2)
$$
is {\em not\/} vanishing by itself, as anticipated in Section \ref{subsec:traces}.


\section{The $n$-particle closed Toda case}\label{FS}

In this section we show that the results obtained for the 4-particle case hold in the general ($n$-particle) case.
\begin{theorem}
\label{teo:toda-n}
Let us consider the PqN structure $(\RR^{2n},\pi,N,\phi)$, where $\pi$ is the canonical Poisson tensor 
and $N$, $\phi$ are given by the obvious generalizations 
of (\ref{qN4},\ref{Phi}). Then 
\begin{enumerate}
\item For all $k\ge 1$, we have that $i_{Y_k}\Omega=0$, where $\Omega=\mathrm{e}^{q_n-q_1}dq_n\wedge dq_1$
and $Y_k=N^{k-1}X_1-X_k$.
\item The functions $I_k=\frac1k\Tr(N^k)$ are in involution.
\end{enumerate}
 \end{theorem}
{\bf Proof.} 

1. {\color{black} Applying $\pi$ to both members of (\ref{ff2}), one easily finds that $NX_l-NX_{l+1}=\pi\,\phi_{l-1}$. Then we have 
\begin{equation}
Y_k=\sum_{l=1}^{k-1}\left(N^{k-l}X_l-N^{k-l-1}X_{l+1}\right)=\sum_{l=1}^{k-1}N^{k-l-1}\left(N X_l-X_{l+1}\right)
=\sum_{l=1}^{k-1}N^{k-l-1}\pi\,\phi_{l-1}\,,
\label{pre-eq:1}
\end{equation}
so that
\begin{equation}
Y_k=\pi\left(\sum_{l=1}^{k-1}(N^*)^{k-l-1}\phi_{l-1}\right)=\pi\left(\sum_{l=0}^{k-2}{(N^*)}^{k-l-2}\phi_l\right)\,.
\label{eq:1}
\end{equation}}
Therefore, the condition $i_{Y_k}\Omega=0$, that is,  $\langle dq_n,Y_k\rangle=\langle dq_1,Y_k\rangle=0$, becomes
\begin{equation}
\sum_{l=0}^{k-2}\brke{\phi_l}{N^{k-l-2}\partial_{p_n}}=\sum_{l=0}^{k-2}\brke{\phi_l}{N^{k-l-2}\partial_{p_1}}=0\,.\label{eq:2}
\end{equation}
Recall now the definition 
\begin{equation}
\label{eq:3}
\brke{\phi_l}{X}=\Tr\left(N^l (i_XT_N)\right)
\end{equation}
of the 1-forms $\phi_l$ and formula (\ref{eq:5}), that is, 
\begin{equation}
i_XT_N=\Delta N (X)\otimes dI_1-\langle dI_1,X\rangle\Delta N+X_1\otimes i_X\Omega\,,\label{eq:5c}
\end{equation}
where $\Delta N$ is given by the obvious generalization of (\ref{delta}),
\begin{equation}
\label{delta-n}
\Delta N=\mathrm{e}^{q_n-q_1}\left(\partial_{p_n}\otimes dq_1-\partial_{p_1}\otimes dq_n\right)\, .
\end{equation}
Then, for all $k\geq 2$ and $l=0,\dots,k-2$, we have that 
\begin{equation}
\label{eq:4}
\begin{split}
&\brke{\phi_l}{N^{k-l-2}\partial_{p_n}}=\Tr\left(N^l(i_{N^{k-l-2}\partial_{p_n}}T_N)\right)\\
&\quad =\Tr \left[N^l\left(\Delta N(N^{k-l-2}\partial_{p_n})\otimes dI_1-\langle dI_1,N^{k-l-2}\partial_{p_n}\rangle\Delta N+
X_1\otimes i_{N^{k-l-2}\partial_{p_n}}\Omega\right)\right]\\
&\quad=\langle dI_1,N^l\Delta N(N^{k-l-2}\partial_{p_n})\rangle-\langle dI_1,N^{k-l-2}\partial_{p_n}\rangle\Tr(N^l\Delta N)+\Omega(N^{k-l-2}\partial_{p_n},N^lX_1)\\
&\quad=2\Omega(N^{k-l-2}\partial_{p_n},N^lX_1)-\langle dI_1,N^{k-l-2}\partial_{p_n}\rangle\Tr(N^l\Delta N),
\end{split}
\end{equation}
where the last equality follows from the identity $\Delta N=-\pi\, \Omega^\flat$. 

Let us compute the three terms appearing in (\ref{eq:4}):
\begin{enumerate}
\item[(i)]
$\Omega(N^{k-l-2}\partial_{p_n},N^lX_1)=
e^{q_n-q_1}\left[\langle dq_n,N^{k-l-2}\partial_{p_n}\rangle\langle dq_1, N^lX_1\rangle-\langle dq_1,N^{k-l-2}\partial_{p_n}\rangle\langle dq_n,N^lX_1\rangle\right].
$
\item[(ii)]
$\langle dI_1,N^{k-l-2}\partial_{p_n}\rangle=-\langle dI_1,N^{k-l-2}(\pi dq_n)\rangle=\langle dq_n,N^{k-l-2}X_1\rangle.$\
\item[(iii)]
$\Tr (N^l\Delta N)
=\langle dq_1,N^l\Delta N(\partial_{q_1})\rangle+\langle dq_n,N^l\Delta N(\partial_{q_n})\rangle 
=e^{q_n-q_1}\big(\langle dq_n,N^l\partial_{p_1}\rangle-\langle dq_1,N^l\partial_{p_n}\rangle\big)
=-2e^{q_n-q_1}\langle dq_1,N^l\partial_{p_n}\rangle.$
\end{enumerate}
Then we proved that
\begin{eqnarray*}
\brke{\phi_l}{N^{k-l-2}\partial_{p_n}}&=&2e^{q_n-q_1}\big[\langle dq_n,N^{k-l-2}\partial_{p_n}\rangle\langle dq_1,N^lX_1\rangle\\
&-&\langle dq_1,N^{k-l-2}\partial_{p_n}\rangle\langle dq_n,N^lX_1\rangle\\
&+&\langle dq_1,N^{l}\partial_{p_n}\rangle\langle dq_n, N^{k-l-2}X_1\rangle].
\end{eqnarray*}
It follows that, for all $k\geq 2$,
\begin{equation}
\begin{split}
\langle dq_n,Y_k\rangle &=\brke{dq_n}{\pi
\sum_{l=0}^{k-2}(N^*)^{k-l-2}\phi_l
}= \sum_{l=0}^{k-2}\brke{\phi_l}{N^{k-l-2}\partial_{p_n}}\\
&=2e^{q_n-q_1}\sum_{l=0}^{k-2} 
\langle dq_n,N^{k-l-2}\partial_{p_n}\rangle  \langle dq_1,N^l X_1\rangle \,,
\end{split}
\end{equation}
proving that if $\langle dq_n,N^j\partial_{p_n}\rangle=0$ for all $j\geq 1$, then  $\langle d q_n,Y_k\rangle=0$ for all $k\ge 1$.
A similar computation shows that $\langle dq_1, Y_k\rangle=0$ is implied by $\langle dq_1,N^{j}\partial_{p_1}\rangle=0$. Hence we are left with proving that the entries 
$(1,n+1)$ and $(n,2n)$ of $N^k$ vanish for all $k\geq 1$. But this follows from the fact that the $n\times n$ block in the upper right corner of $N^k$ is skewsymmetric, 
since $N^k\pi=\pi\, {N^*}^k$.

2. It suffices to show that the additional term, appearing in (\ref{recadd}), to the usual Lenard-Magri recursion relations for the Poisson 
brackets between the traces of the powers of $N$ 
vanishes. Actually, this additional term is 
 \begin{equation}
\Delta_{j,k-1}=-\brke{ \phi_{j-1}}{\pi\, dI_{k-1}}-\brke{\phi_{k-2}}{ \pi\, dI_j}
\end{equation}  
and it reads, thanks to (the generalization to arbitrary $n$ of) equation (\ref{eqtr}),
\begin{equation}\label{qf2}
\Delta_{j,k-1}=-2\Omega(X_{k-1},N^{j-1}X_1)-2\Omega(X_{j},N^{k-2}X_1)\,.
\end{equation}
Now, thanks to the first part of this theorem, we can substitute $N^{i-1}X_1$ with $X_{i}$ in the previous formula for $\Delta_{j,k-1}$, showing that it vanishes.
Hence we obtain that the Lenard-Magri recursion relations (\ref{LMrec}) hold also in this case, leading to the involutivity of the $I_k$.
\hfill$\square$ \medskip

We notice that in many points of the previous proof (see, e.g., item (iii)) very peculiar properties of the tensor field $\Delta N$ have been exploited.


\section{A relation between PN and PqN manifolds}
\label{igor-connection}

In this section we present a general result concerning the connection between PN and PqN structures. When applied to the PqN manifold we have studied in the previous section, this result allows us to establish a relation between the geometrical structures of the closed and open Toda lattices. 

First of all, we recall that, given a tensor field $N:T\M\to T\M$, the usual Cartan differential can be modified as follows,
\begin{equation}
\label{eq:dN}
\begin{split}
(d_N\alpha)(X_0,\dots,X_q)&=
\sum_{j=0}^q(-1)^j L_{NX_j}\left(\alpha(X_0,\dots,\hat{X}_j,\dots,X_q)\right)\\&
+\sum_{i<j}(-1)^{i+j}\alpha([X_i,X_j]_N,X_0,\dots,\hat{X}_i,\dots,\hat{X}_j,\dots,X_q)\,,
\end{split}
\end{equation}
where $\alpha$ is a $q$-form, the $X_i$ are vector fields,
and 
$[X,Y]_N=[NX,Y]+[X,NY]-N[X,Y]$.  Note that 
$d_Nf=N^* df$ for all $f\in C^\infty(\M)$. Moreover, 
\begin{equation}
\label{eq:dNd}
d_N=i_N\circ d-d\circ i_N\,,
\end{equation}
where $i_N$ is given by (\ref{iNalpha}), and consequently $d\circ d_N+d_N\circ d=0$. 
Finally, $d_N^2 =0$ if and only if the torsion of $N$ vanishes.

We also remind that one can define a Lie bracket between the 1-forms on a Poisson manifold $(\M,\pi)$ as
\begin{equation}
\label{eq:liealgpi}
[\alpha,\beta]_\pi=L_{\pi\alpha}\beta-L_{\pi\beta}\alpha-d\langle\beta,\pi\alpha\rangle\,,
\end{equation}
and that this Lie bracket 
can be uniquely extended to all forms on $\M$ in such a way that
\begin{itemize}
\item[(K1)] $[\eta,\eta']_\pi=-(-1)^{(q-1)(q'-1)}[\eta',\eta]_\pi$ if $\eta$ is a $q$-form and $\eta'$ is a $q'$-form;
\item[(K2)] $[\alpha,f]_\pi=i_{\pi df}\,\alpha=\langle\alpha,\pi df\rangle
$ for all $f\in C^\infty(M)$ and for all 1-forms $\alpha$;
\item[(K3)] if $\eta$ is a $q$-form, then $[\eta,\cdot]_\pi$ 
is a derivation of degree $q-1$ of the wedge product, that is,
\begin{equation}
\label{deriv-koszul}
[\eta,\eta'\wedge\eta'']_\pi=[\eta,\eta']_\pi\wedge\eta''+(-1)^{(q-1)q'}\eta'\wedge[\eta,\eta'']_\pi
\end{equation}
if $\eta'$ is a $q'$-form and $\eta''$ is any differential form.
\end{itemize}
This extension is a {\it graded\/} Lie bracket, in the sense that (besides (K1)) the graded Jacobi identity holds:
\begin{equation}
\label{graded-jacobi}
(-1)^{(q_1-1)(q_3-1)}[\eta_1,[\eta_2,\eta_3]_\pi]_\pi+(-1)^{(q_2-1)(q_1-1)}[\eta_2,[\eta_3,\eta_1]_\pi]_\pi+(-1)^{(q_3-1)(q_2-1)}[\eta_3,[\eta_1,\eta_2]_\pi]_\pi=0
\end{equation}
if $q_i$ is the degree of $\eta_i$.
It is sometimes called the Koszul bracket --- see, e.g., \cite{FiorenzaManetti2012} and references therein.

It was proved in \cite{YKS96} that the compatibility conditions (\ref{N-P-compatible}) between a Poisson tensor $\pi$ and a tensor field $N:T\M\to T\M$ hold if and only if 
$d_N$ is a derivation of $[\cdot,\cdot]_\pi$, 
that is,
\begin{equation}
\label{deriv-wedge}
d_N[\eta,\eta']_\pi=[d_N\eta,\eta']_\pi+(-1)^{(q-1)}[\eta,d_N\eta']_\pi
\end{equation}
if $\eta$ is a $q$-form and $\eta'$ is any differential form. In particular, taking $N=Id$, one has that the Cartan differential $d$ is always a derivation of  
$[\cdot,\cdot]_\pi$. Moreover,
if $\phi$ is any 3-form,
\begin{equation}
\label{iff:SX}
d_{N}^2=[\phi,\cdot]_\pi\quad\mbox{if and only if}\quad 
\left\{\begin{array}{l}
T_{N}(X,Y)=\pi\left(i_{X\wedge Y}\phi\right)\quad\mbox{ for all vector fields $X,Y$}\\
\noalign{\medskip}
i_{(\pi\alpha)\wedge(\pi\beta)\wedge(\pi\gamma)}(d\phi)=0\quad\mbox{ for all 1-forms $\alpha,\beta,\gamma$,}
\end{array}\right.
\end{equation}
see \cite{SX}. We are now ready to state
\begin{theorem}
\label{thm:NEW}
Suppose that $(\M,\pi,\phi,N)$ is a PqN manifold and that there exists a closed 2-form $\omega$ such that 
\begin{equation}
\label{Mau-Car}
d_N\omega+\frac{1}{2}[\omega,\omega]_\pi=-\phi\,.
\end{equation}
If $N'=N-\pi\,\omega^{\flat}$, then 
$(\M,\pi,N')$ is a PN manifold. 
\end{theorem}
{\bf Proof.}
First of all we show that $d_{\pi\,\omega^{\flat}}=-[\omega,\cdot]_\pi$. This follows from the fact that both are derivations (with respect to the wedge product) 
anti-commuting with $d$, and they coincide on functions. Indeed, for all $f\in C^\infty(\M)$, 
\[
d_{\pi\,
\omega^{\flat}}f=(\pi\,
\omega^{\flat})^*df=(\omega^{\flat}
\pi)df=i_{\pi df}\,\omega=-[\omega,f]_\pi,
\]
where the last equality holds for every 2-form $\omega$ and can be easily checked to be a consequence of (K2) and (K3).

Hence $d_{N'}=d_N-d_{\pi\, 
\omega^{\flat}}=d_N+[\omega,\cdot]_\pi$ 
is a derivation of $[\cdot,\cdot]_\pi$ (since $\pi$ and $N$ are compatible and $[\cdot,\cdot]_\pi$ satisfies (\ref{graded-jacobi})), so that 
$\pi$ and $N'$ are compatible too.

Finally, equivalence (\ref{iff:SX}) and formula (\ref{Mau-Car}) imply that $d_{N'}^2=0$, meaning that the torsion of $N'$ vanishes. We conclude that $(\M,\pi,N')$ is a PN manifold. 
\hfill$\square$ \medskip

In the terminology of \cite{ILX}, Theorem \ref{thm:NEW} describes how to deform a quasi-Lie bialgebroid into a Lie bialgebroid by means of the so called {\it twist}. 

\begin{rem}
\label{ex:toda}
Twisting the PqN structure of the Toda closed system with 
\begin{equation}
\label{eq:todatwist}
{\color{black}\omega=-\Omega=-e^{q_n-q_1}dq_n\wedge dq_1=-d\left(e^{q_n-q_1} dq_1\right)\,,}
\end{equation}
one obtains the the PN structure of Toda open system. In fact, 
$[\omega,\omega]_\pi={\color{black}[\Omega,\Omega]_\pi=}0$ and $d_N\omega={\color{black}-d_N\Omega=}-\phi$, 
so that (\ref{Mau-Car}) is satisfied. Moreover, using the notations of Section \ref{toda-4} and and \ref{FS}, we have 
that $N'=N_{(O)}=N-\Delta N$, where $\Delta N=\pi\,\omega^{\flat}{\color{black}=-\pi\,\Omega^{\flat}}$.
\end{rem}

A kind of converse of Theorem \ref{thm:NEW} is given by
\begin{theorem}
Let $(M,\pi,N)$ be a PN manifold. Then:
\begin{enumerate}
\item
For every closed 2-form $\omega$ such that 
\begin{equation}
d_N\omega+\frac{1}{2}[\omega,\omega]_\pi=0\,,\label{eq:MComega}
\end{equation}
defining $N'=N-\pi\,
\omega^\flat$, we have that $(M,\pi,N')$ is a PN manifold.
\item 
Let $\omega$ 
be a closed 2-form such that 
\begin{equation}
\label{eq:cond}
[d_N\omega,\omega]_\pi=0\,. 
\end{equation}
If 
\begin{equation}
\label{eq:condiphi}
\phi=d_N\omega+\frac{1}{2}[\omega,\omega]_\pi
\end{equation} 
and $N'=N-\pi\,\omega^\flat$, then $(M,\pi,N',\phi)$ is a PqN manifold.
\end{enumerate}
\end{theorem}
{\bf Proof.}
Part 1 is Theorem \ref{thm:NEW} with $\phi=0$.
As far as part 2 
is concerned, note that condition \eqref{eq:cond} guarantees that the 3-form $\phi$ defined by \eqref{eq:condiphi} satisfies $d_N\phi=0$ and $d\phi=0$. Thanks to (\ref{eq:dNd}), it follows that $i_N\phi$ is closed. Since $d_{N'}=d_N-d_{\pi\, 
\omega^{\flat}}=d_N+[\omega,\cdot]_\pi$, the compatibility between $\pi$ and $N'$ can be shown as in the proof of Theorem \ref{thm:NEW}.
Finally, using \eqref{eq:condiphi} and $d_N^2=0$, we can prove that $d_{N'}^2=[\phi,\cdot]_\pi$. To conclude, it suffices to use equivalence (\ref{iff:SX}).
\hfill$\square$\medskip

\begin{rem} We conclude this section with a couple of remarks.
\begin{enumerate}
\item In the notations of Sections \ref{toda-4} and \ref{FS}, starting from the open Toda system, we can consider the closed 2-form 
$\omega=\Omega=e^{q_n-q_1}dq_n\wedge dq_1$.  
{\color{black} As we have seen in Remark \ref{ex:toda}, it satisfies $[\Omega, \Omega]_\pi = 0$. One can also show that}
$d_{N_{(O)}}\Omega=\phi$, so that {\color{black} (\ref{eq:condiphi}) is fulfilled and} condition \eqref{eq:cond} becomes $[\phi,\Omega]_\pi=0$, {\color{black} which is a direct consequence of 
$[\Omega, \Omega]_\pi = 0$.}
Note that this computation involves the (torsionless) tensor $N_{(O)}$ of the (PN structure of the) Toda open system, while the analogue computation in Remark \ref{ex:toda} involves the tensor $N$ of the (PqN structure of the) Toda closed system, whose torsion does not vanish.
\item \textcolor{black}{To the best of our knowledge equation \eqref{eq:MComega} was first introduced and studied by Liu, Weinstein and Xu in their work on the theory of Manin triples for Lie algebroids, see Section 6 of \cite{LWX}. These authors, starting from a Poisson manifold $(\mathcal M,\pi)$ and the corresponding \emph{standard} Courant algebroid structure on $T^\ast\mathcal M\oplus T\mathcal M$, showed that for $N=Id$ every solution of \eqref{eq:MComega} defines a Dirac subbundle 
$\Gamma_\omega\subset T^\ast\mathcal M\oplus T\mathcal M$ transversal to $T^\ast\mathcal M$. Moreover they proved that every solution of
\begin{equation}
d\omega=0\quad\text{and}\quad[\omega,\omega]_\pi=0\label{complementary}
\end{equation}
defines a new Poisson structure $\pi'$ on $\mathcal M$ compatible with $\pi$ and induced by a torsionless operator, defining in this way a Poisson-Nijenhuis structure on $\mathcal M$. It is worth to mention that the second equation in \eqref{complementary} was studied in depth by Vaisman in \cite{Complementary}, where its solutions were named complementary $2$-forms of the (underlying) Poisson structure.}
\end{enumerate}
\end{rem}


\end{document}